# Physics experiments using simultaneously more than one smartphone sensors

**M MONTEIRO[1], C STARI[2], C CABEZA[2] and A C MARTÍ[2]**
[1]*Universidad ORT Uruguay*
[2]*Universidad de la República, Uruguay*

**Abstract.** In the last years, numerous Physics experiments using smartphone sensors have been reported in the literature. In this presentation we focus on a less-explored feature of the smartphones: the possibility of using (measure and register data) simultaneously with more than one sensor. To illustrate, in the field of mechanics simultaneous use of the accelerometer and gyroscope (angular velocity sensor) or in optics experiments synchronous use of the ambient light and orientation sensors have been proposed. Indeed, this is a characteristic that simplifies experimental setups allowing to see through the physics concepts and, last but not least, reducing the costs.

## 1. Introduction

Recently, numerous smartphone-based physics experiments have been proposed in the literature [1,18]. These experiments take advantage of the built-in smartphone sensors as the accelerometer, gyroscope (angular velocity sensor), magnetometer, proximeter, luxometer (ambient light), pressure (barometer) among others. Usually, only one sensor is used in each experiment. Remarkably, smartphones also gives us the ability to measure *simultaneously* with various sensors. This is a great benefit since it allows to perform a great deal of experiments, even outdoors, avoiding the dependence on delicate and expensive instruments. As we consider that this capability has not been fully exploited, in this presentation we discuss some experiments involving the use of more-than-one smartphone sensor and, at the end, present some perspectives. In table I we list the most commonly used sensors and the abbreviations used in this text.

**Table 1.** Most commonly used sensors.

| Abbreviations | Sensor |
|---|---|
| Ac | Accelerometer |
| Gy | Gyroscope |
| Mi | Microphone |
| Ca | Camera |
| Li | Light |
| Pr | Pressure |
| Or | Orientation |
| GPS | GPS |
| Ma | Magnetometer |

## 2. Simultaneous use of more than one smartphone sensors

As far as we know, in the first Physics experiment in which more than one sensor was used a smartphone was placed on the floor of a merry-go-round, as shown in figure 1, at different distances from the rotation shaft [3]. In this experiment, the accelerometer and angular velocity (gyroscope) sensors were used simultaneously to obtain the centripetal acceleration, $a_c$, and the angular velocity, $\omega$, and verify the elementary relationship, $a_c = R\omega^2$, between those variables and by means of a linear regression obtain the rotation radius.

Table 2. A brief summary of the use of the sensors.

| References | Sensors involved | Experimental use |
|---|---|---|
| [3] | **Ac + Gy** | Verify fundamental relationship between centripetal acceleration and angular velocity |
| [4] | **Ac + Gy** | Obtain mechanical energy (translational and rotational) |
| [5] | **Ac + Gy** | Obtain trajectories in phase space |
| [6] | **Ac + Gy+Ca** | Overcome limitations from the equivalence principle |
| [12] | **Ac + Gy** | Obtain Coriolis acceleration |
| [13] | **Ac + Gy + Pr** | Several experiments in an amusement park |
| [14] | **Ac + GPS** | Determination of the drag coefficients of various vehicles |
| [10] | **Ac + Ma** | Obtain spatial dependence of magnetic field |
| [7] | **Ac + Pr** | Comparison of results obtained with different sensors |
| [11] | **Ca + Gy** | Relate fluid surface's shape to angular velocity |
| [8] | **GPS + Pr** | Obtain spatial dependence of atmospheric variables |
| [9] | **Li + Or** | Obtain angular dependence of light intensity |
| [15] | **Li + Or** | Obtain angular dependence of solar irradiance |
| [16] | **Mi + Or** | Visualizing sound directivity |
| [17] | **Ac + Gy** | Several experiments in an amusement park and Coriolis |

The physical pendulum, one of the most paradigmatical mechanical system, was also studied using simultaneously the accelerometer and gyroscope [4-6]. In this setup, a smartphone affixed to a bicycle wheel was subject to both rotational and oscillatory motion. Thanks to the smartphone's capabilities two acceleration components and the angular velocity can be readily obtained. Several activities can be proposed from this experiment. A secondary or high-school lab can be focused on the rotational kinetic energy and the characteristics of the oscillatory motion [4]. In intermediate contexts interesting questions about the equivalence principle can be raised [5], while in advances laboratories, in this relatively simple system with one degree of freedom, a generalized coordinate and the conjugate momentum can be determined, enabling the representation of trajectories in the phase space [6]. In this way, this latter, rather abstract, concept is rendered more tangible as the students can visualize trajectories directly from the experimental data provided by the sensors.

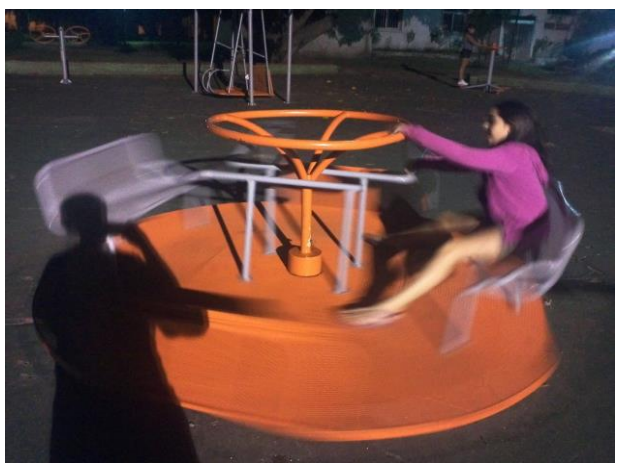

**Figure 1.** A smartphone located in a merry-go-round is employed to simultaneously measure the angular velocity and the centripetal acceleration.

The accelerometer can also be employed in conjunction with the pressure sensor to obtain (and corroborate the coherence of) the vertical component of the velocity of elevators, pedestrians climbing stairs, and flying unmanned aerial vehicles [7]. This is an example of outreach activity that can be performed with student or a general audience. It is shown in this reference that the pressure sensor outperforms the accelerometer and GPS in several contexts.

In another experiment, the pressure sensor and the GPS were used in synchrony to find the relationship between atmospheric pressure and altitude [8] and gain insight into the characteristics of the inner atmosphere. In this case the setup is a bit more complex because the smartphone is mounted on a quadcopter. Both the altitude and the pressure are obtained using the built-in sensors. The results can be compared with reference values and other simple approximations as the isothermal and constant density atmospheres.

Optics also offers the possibility of doing smartphone experiments. Recently, the advantage of the capabilities of a smartphone to verify the Malus' law was proposed [9]. In this case, the intensity of polarized light from a computer monitor is measured by means of the luxometer (or ambient light sensor) with a small polarizer attached to it while the angle between the polarization and the polarizer is measured using of the orientation sensor. The simultaneous use of these two sensors allows us to simplify the experimental setup and complete a set of measures in just a few minutes. An exemplification of the results is shown in figure 2.

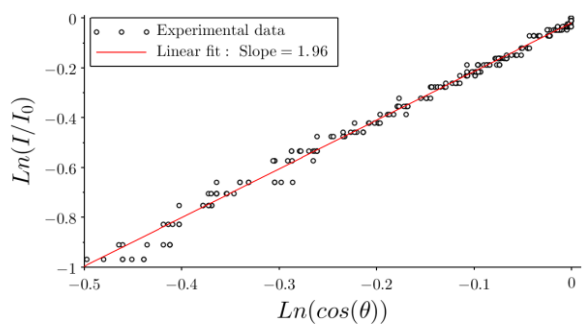

**Figure 2.** Malus law and light polarization experiment. Dependence of the light intensity as a function of the angle. For further information see [9].

In a different mixture of sensors the magnetometer and accelerometer were also used in synchrony [10] to obtain the spatial dependence of magnetic fields in simple configurations. In the simplest version, a smartphone is mounted on a track whose direction coincides with the axis of a coil. In the experimental setup, shown in figure 3, the smartphone is gently accelerated and both the distance (integrated numerically from the acceleration values) and the magnetic field are simultaneously obtained.

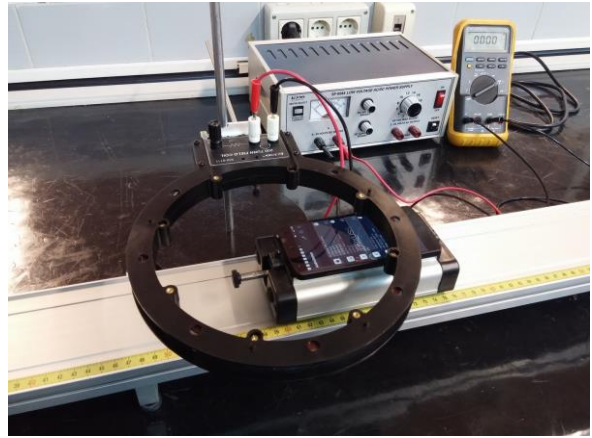

**Figure 3.** Experimental setup similar to that of Ref.[10] in which the spatial dependence of a magnetic field is obtained. In this experiment, numerically integrating data provided by the accelerometer the position is obtained.

The surface of a liquid in a rotating frame is studied in [11]. In this experiment, a fluid in a rectangular container with a small width is placed on rotating table. A smartphone fixed to the rotating frame simultaneously records the fluid surface with the camera and also, thanks to the built-in gyroscope, the angular velocity. When the table starts rotating the surface evolves. Using video analysis the surface's shape is obtained and the concavity related to the angular velocity. An elusive magnitude, the Coriolis acceleration, is measured in an undergraduate experiment proposed in [12]. The simultaneous use of the gyroscope and accelerometer helps to verify the dependence of Coriolis acceleration on the angular velocity of a rotating track and the speed of the sliding smartphone. In [19] an amusement park is the scenario for an experiment using both the accelerometer and the gyroscope but involving three axes of rotation.

Two experiments [13,14] stand out in the field of aerodynamics. In the first [13], the experiment takes place in a sky roller where forces, torque and angular velocities during different parts of the ride were analyzed. It is worth noting that in this case, three sensors (accelerometer, gyroscope and pressure) are involved. As far as we know, this is the only experiment which uses more than two different sensors. In the second, as another practical application, the measurement of the drag coefficient of a commercial automobile by using the acceleration sensor in conjunction with the GPS is discussed in [14].

Concerning other less used sensors, a non-expensive setup is introduced in [15], to measure the solar irradiance using simultaneously the ambient light and orientation sensors. This magnitude is relevant in photovoltaic systems and usually requires sophisticated experimental setup. A promising approach is proposed in [16], in which, using microphones and orientation sensor students can gain insight into the properties of sound directivity, interference and other acoustical phenomena.

**Table 3.** Simultaneous use of two sensors

|               | Ac.       | Gy.                    | Mi.      | Ca.      | Li.     | Pr.      | Or.     | GPS   | Ma.  |
|---------------|-----------|------------------------|----------|----------|---------|----------|---------|-------|------|
| Accelerometer |           | [3, 4, 5, 6, 12, 13,19]|          |          |         | [7, 13]  |         | [14]  | [10] |
| Gyroscope     | [3, 4, 5, 6, 12, 13,19] |         |          | [6, 11]  |         |          |         |       |      |
| Microphone    |           |                        |          |          |         |          | [16]    |       |      |
| Camera        |           | [6, 11]                |          |          |         |          |         |       |      |
| Light         |           |                        |          |          |         |          | [9, 15] |       |      |
| Pressure      | [7, 14]   |                        |          |          |         |          |         | [8]   |      |
| Orientation   |           |                        | [16]     |          | [9, 15] |          |         |       |      |
| GPS           | [14]      |                        |          |          |         |          | [8]     |       |      |
| Magnetometer  | [10]      |                        |          |          |         |          |         |       |      |

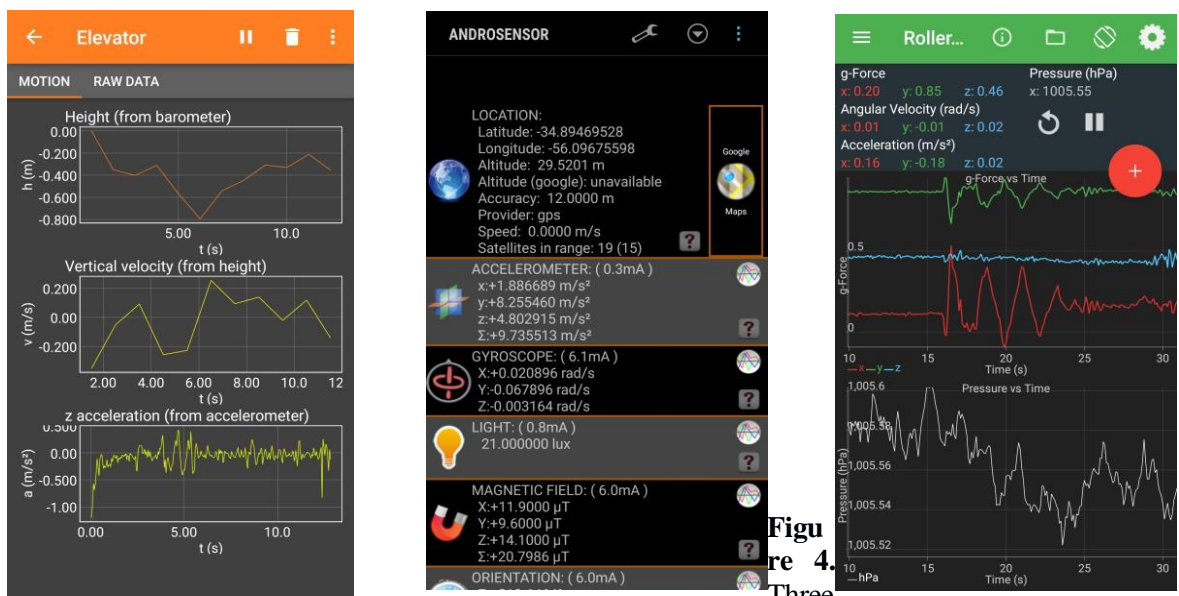

**Figure 4.** Screen-shots of the applications that allow registering data with multiple sensors: Phyphox, Androsensor and Physics Toolbox Suite.

Finally, let us mention the *apps* employed in various of the experiments. These are Phyphox, Physics Toolbox Suite and Androsensor and they are available for the main operating systems in the usual websites. All of them allow to export data, using CSV file format, to a tablet, notebook or desktop computer. In figure 4 we show examples of the experiments presented.

### 3. Conclusion and perspectives
The overall panorama of simultaneous use of sensors is depicted in tables 2 and 3. A brief summary of the use of the sensors is depicted in table 2. As appreciated in the discussed examples the synchronous use of several smartphone sensors opens the possibility of numerous physics experiments. Most modern smartphones present around a dozen of sensors. From simple arithmetic, using only two sensors, there are about half a hundred of combinations of sensors. In table 3 we represent with red boxes the two-sensor combinations that have been employed in at least one experiment. White boxes correspond to places in which there are, until now, not proposals. So, let our imagination soar and devise new experiments.